\documentclass[aps,twocolumn]{revtex4}%
\usepackage{amsfonts}
\usepackage{amsmath}
\usepackage{amssymb}
\usepackage{hyperref}
\usepackage{color}
\usepackage{graphicx}%

\begin{document}
\title{Magnetization of Two Dimensional
Heavy Holes with Boundaries in a Perpendicular Magnetic Field}
\author{Cheng Fang$^{a,b,c}$}
\author{Zhigang Wang$^{c}$}
\author{Shu-Shen Li$^{a}$}
\author{Ping Zhang$^{c,d}$}
\thanks{Corresponding author. E-mail:zhang\_ping@iapcm.ac.cn}

\affiliation{$^{\left.a\right)}$State Key Laboratory for
Superlattices and Microstructures, Institute of Semiconductors,
Chinese Academy of Sciences, P.O. Box 912, Beijing 100083, P.R.
China
\\$^{\left.b\right)}$Physics Department, East China Institute of
Technology, Fuzhou,
Jiangxi, 344000, P.R. China\\
$^{\left.c\right)}$LCP, Institute of Applied Physics and
Computational Mathematics, P.O. Box 8009, Beijing 100088, P.R.
China\\
$^{\left.d\right)}$Center for Applied Physics and Technology, Peking
University, Beijing 100871, People's Republic of China}
\begin{abstract}
The magnetization of heavy holes in III-V semiconductor quantum
wells with Rashba spin-orbit coupling (SOC) in an external
perpendicular magnetic field is theoretically studied. We
concentrate on the effects on the magnetization induced by the
system boundary, the Rashba SOC and the temperature. It is found
that the sawtooth-like de Haas--van Alphen (dHvA) oscillations of
the magnetization will change dramatically in the presence of such
three factors. Especially, the effects of the edge states and Rashba
SOC on the magnetization are more evident when the magnetic field is
more small. The oscillation center will shift when the boundary
effect is considered and the Rashba SOC will bring beating patterns
to the dHvA oscillations. These effects on the dHvA oscillations are
preferred to be observed at low temperature. With increasing the
temperature, the dHvA oscillations turn to be blurred and eventually
disappear.

\noindent%
\end{abstract}
\maketitle

\section{Introduction}

The physics of low dimensional electron systems in high magnetic
fields is one of the most important subjects of semiconductor
physics. Recently, great attentions are taken to the magnetic
properties of two dimensional (2D) systems in a strong perpendicular
magnetic field both experimentally and
theoretically\cite{PRB67.155329,PRB68.193308,PhysRevB.61.4835,PhysRevB.62.15367,PhysRevLett.86.4644}.
Such systems are candidates for next-generation spintronic
devices\cite{H.Ohno08141998,S.A.Wolf11162001,J.A.Gupta06292001}
because of their very high electron mobility compared with silicon.
In the absence of disorder and interactions, due to the Landau
energy quantization in an external magnetic field, the magnetization
is predicted to oscillate periodically in a sawtooth pattern as a
function of $B$ or filling factor $\nu$, which is the famous de
Haas-- van Alphen (dHvA)
oscillation\cite{PeierlsZPhys1933,PhysRevLett.51.1700,JLowTempPhys56.417}.
The dHvA oscillation is a powerful tool used extensively to
determine the electronic properties of bulk semiconductors and
metals, such as Fermi surface geometry and  effective masses. It has
been evident now that the magnetization, which turns to be zero for
classical electrons, appears due to the presence of the sample
boundary\cite{PRB59.7305}, the spin orbital coupling (SOC) or Zeeman
splitting. Once a perpendicular magnetic field was applied, these
factors would change the magnetization dramatically. Firstly, in a
finite size sample with boundaries the edge states play important
roles in quantum transport in high magnetic fields, such as quantum
Hall effect or magneto-tunneling effect. From the semiclassical
point of view when electrons are located near the edge of the
samples, cyclotron motion cannot fulfill the entire cyclotron orbit
if the distance between the center of the cyclotron orbit and the
edge is smaller than the radius of the cyclotron motion. In such a
case, the electron orbit is bounced at the edge plane, and the
bounced electrons conduct a one-dimensional motion as a whole being
repeatedly reflected at the edge plane, resulting a skipping orbit.
The edge states with a skipping orbit are different from those in
the interior of the sample where the cyclotron orbit is completed.
Although these edge states are localized in the vicinity of the
confining walls, the measurable bulk quantities are completely
modified by the edge states, certainly including the magnetization.
Secondly, the Rashba SOC, competed with the Landau splitting and the
Zeeman splitting, will modulate the energy spectrum. As a result,
the Rashba SOC will have non-neglectable influence on the dHvA
oscillations of the magnetization.

However, to our knowledge there are no detailed treatments on the
influence of edge states and SOC on the magnetization in 2D systems,
except for the 2D electron systems much recently studied by our
group\cite{wang}. In this paper, we study systematically the
thermodynamic magnetization of heavy holes in III-V semiconductor
quantum wells with boundaries in the presence of the Rashba SOC due
to structure-inversion asymmetry, with an external magnetic field
applied perpendicularly. We will show that the dHvA oscillations of
the magnetization change dramatically due to the presence of SOC and
system boundaries. The edge states lead to the change of both the
center and the amplitude of the sawtoothlike dHvA oscillations of
the magnetization. The SOC mixes the spin-up and spin-down states of
neighboring Landau Levels into two unequally spaced energy branches,
which further changes the well-defined sawtoothlike dHvA
oscillations of the magnetization. These effects on the
magnetization can be observed at low temperature in experiment. With
increasing the temperature, the dHvA oscillations of the
magnetization turn to be blurred and eventually disappear.

This paper is organized as follows:
Sec.\ref{sec:The-Quantum-Solution} focuses on the quantum mechanical
solution of the heavy-holes in III-V semiconductor quantum wells.
Using numerical diagonalization of the Hamiltonian in a truncated
Hilbert space we calculate the energy spectrum.
Sec.\ref{sec:OM_formula} gives the main results of this paper on the
magnetization and makes a detailed discussion. The last section
gives a short summary.

\section{Energy spectrum of the heavy holes\label{sec:The-Quantum-Solution}}

\begin{figure}[ptb]
 \includegraphics[scale=0.8]{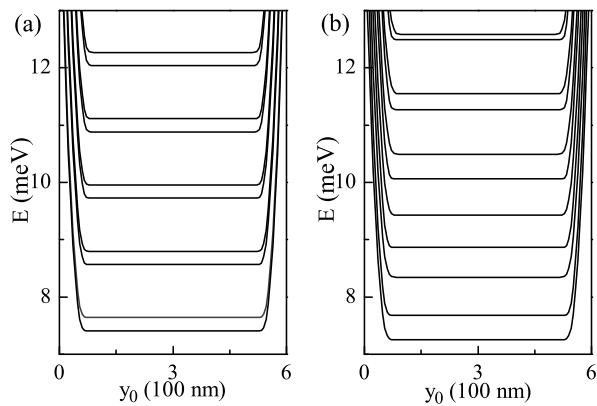}

\caption{The energy spectrum in units of meV versus the guiding
center $y_{0}$ without (a) and with (b) the Rashba SOC,
respectively, for the 2D hole systems. In both figures, the
effective mass $m^{\ast}$=$0.27m_{e}$, the system size
$L$=$600\,$nm,$\;$the hole number density $N_{h}$=$0.01$ nm$^{-2}$,
the magnetic field $B$=$3$T, and the $g$-factor $g_{s}$=$-0.44$.
$\;$ The Rashba SOC strength in (b) is set as
$\alpha$=$40\,$meV\,nm$^{3}$. Clearly from (b) one can see that the
introduced Rashba SOC mixes the the eigenstates of the spin operator
$s_{z}$. }

\label{energy}
\end{figure}

Now we consider a 2D hole gas system in which a Rashba SOC arises
from the quantum well asymmetry in the growth ($z$). The Hamiltonian
for heavy holes in III-V semiconductor quantum wells within a
perpendicular magnetic field $\mathbf{B}=-B\hat{e}_{z}$, taking into
account the kinetic energy and the Zeeman magnetic field, can be
written as\begin{equation}
H=\frac{\Pi^{2}}{2m^{\ast}}+i\frac{\alpha}{2\hbar^{3}}\left(\Pi_{-}^{3}\sigma_{+}-\Pi_{+}^{3}\sigma_{-}\right)-\frac{3}{2}g_{s}
\mu_{B}\mathbf{B}\sigma_{z}+V\left(y\right),\label{h}\end{equation}
 where $m^{\ast}$ being band effective mass, $\Pi_{\pm}=\Pi_{x}\pm i\Pi_{y}$
with $\Pi_{\eta}=\Pi_{\eta}+\left(e/c\right)A_{\eta}$, $\Pi_{\eta}$
and $A_{\eta}$ denoting the $\eta$ component of the momentum and
vector potential, respectively.
$\mathbf{B}$=$\nabla\times\mathbf{A}$. The Pauli matrices,
$\sigma_{\pm}=\sigma_{x}\pm i\sigma_{y}$, operate on the total
angular momentum states with spin projection $\pm3/2$ along the
growth direction; in this sense they represent a pseudospin degree
of freedom rather than a genuine spin
$1/2$\cite{PhysRevB.71.085308}. $\alpha$ is Rashba SOC coefficient
due to structure inversion asymmetry across the quantum well grown
along the $[001]$-direction chosen as the $z$-axis. For a
symmetrically grown quantum well, the coefficient $\alpha$ is
essentially proportional to an electric field applied across the
well and therefore experimentally
tunable\cite{PalaEPL2004,pala:045304,PhysRevB.71.085308,winkler:4245,PhysRevB.65.155303,Sov.Phys.Semi26.73}.
$g_{s}$ is the gyromagnetic factor. The last term $V\left(y\right)$
is the lateral confining potential. For simplicity, a hard wall
potential that confines the holes in the transverse $y$ direction is
used,\begin{equation} V\left(y\right)=\begin{cases}
0 & 0\leqslant y\leqslant L\\
\infty & \text{otherwise}\end{cases}.\label{confined
potential}\end{equation}
 It is convenient to use the Laudau gauge $\mathbf{A}=\left(yB,0,0\right)$
and to write the wave function in the form \begin{equation}
\Psi\left(x,y\right)=e^{ikx}\psi\left(y\right)\label{wxy}\end{equation}
 with the function $\psi\left(y\right)$ expanded in the basis set
of the infinite potential well,\begin{equation}
\psi\left(y\right)=\sqrt{\frac{2}{L}}\sum_{n}\sin\left(\frac{\pi
ny}{L}\right)\left(\begin{array}{c}
a_{n}\\
b_{n}\end{array}\right).\label{wy}\end{equation}
 Introducing the cyclotron center $y_{0}=-\frac{\hbar c}{eB}k=-l_{b}^{2}k$,
the magnetic length $l_{b}=\sqrt{\hbar c/eB}$ and the cyclotron
frequency $\omega_{c}=eB/m^{\ast}c$, the Schr\"{o}dinger equation
$H\Psi\left(x,y\right)=E\Psi\left(x,y\right)$ leads to the following
equations for spinors:
\begin{widetext}
\begin{equation} \sum_{l}\left\{
\left[M_{ln}-\hbar\omega_{c}\left(\frac{3}{2}g\sigma_{z}+\varepsilon\right)\delta_{ln}\right]+i\left[F_{ln}-G_{ln}\right]
\sigma_{+}-i\left[F_{ln}+G_{ln}\right]\sigma_{-}\right\}
\left(\begin{array}{c}
a_{n}\\
b_{n}\end{array}\right)=0,\label{SpinEQ}\end{equation}

 where

 \begin{align*}
M_{ln} &
=\hbar\omega_{c}\frac{1}{\pi}\int_{0}^{\pi}\sin\left(lt\right)\left[\left(\frac{L}{\pi
l_{b}}\right)^{2}\left(t-t_{0}\right)^{2}-
 \left(\frac{\pi l_{b}}{L}\right)^{2}\partial_{y}^{2}\right]\sin\left(nt\right)dt,\\
F_{ln} &
=\hbar\omega_{c}\frac{\eta}{L}\int_{0}^{\pi}\sin\left(lt\right)\left[3\partial_{t}+3\left(t-t_{0}\right)\partial_{t}^{2}
+
\left(\frac{L}{\pi l_{b}}\right)^{4}\left(t-t_{0}\right)^{3}\right]\sin\left(nt\right)dt,\\
G_{ln} & =\hbar\omega_{c}\left(\frac{L}{\pi
l_{b}}\right)^{2}\frac{\eta}{L}\int_{0}^{\pi}\sin\left(lt\right)\left[3\left(t-t_{0}\right)
+3\left(y-y_{0}\right)^{2}\partial_{t}+\left(\frac{\pi
l_{b}}{L}\right)^{4}\partial_{t}^{3}\right]\sin\left(nt\right)dt,\end{align*}
\end{widetext}
 with $\varepsilon=E/\hbar\omega_{c}$,$\,\ t=\pi y/L$, $\, t_{0}=\pi y_{0}/L$,
$g=m^{\ast}g_{s}/2m_{e}$, and $\eta=m^{\ast}\alpha l_{b}/\hbar^{2}$.
Taking the same method in Ref.\cite{PhysRevB.70.235344}, we solve
the above equations in a truncated Hilbert space disregarding the
highest energy states. Typically we take a matrix Hamiltonian of
dimension of a few hundred. We increase the size of the Hilbert
space by a factor $2$ and find no change in the results presented
below. In all cases the width of the sample $L$ is taken large
enough to have the cyclotron radius $r_{c}$ smaller than $L/2$. The
right and left edge states are then well separated in real space.
For $y_{0}\simeq L/2$ the states are equal to the \emph{bulk}
states, except for exponential corrections. The energy spectrum
reproduces the bulk results without edge states, which is given by
the follows:

\begin{equation}
\frac{E_{n}}{\hbar\omega_{c}}=\left(n-1\right)+\frac{3}{2}\left(1-g\right),\text{
\ \ }\left(n=0,1,2\right)\label{e0}\end{equation}

\begin{eqnarray}
\frac{E_{n}}{\hbar\omega_{c}}&=&\left(n-1\right)+\frac{s}{2}\sqrt{4\gamma^{2}n\left(n-1\right)\left(n-2\right)+\left(3g-3\right)^{2}},
\nonumber\\
&&\hspace{5.5cm}\left (n\geq3\right)\label{e1}
\end{eqnarray}
with $s$=$\pm1$. The dimensionless parameter is defined by
$\gamma=2\left(\alpha m^{\ast}/\hbar^{2}\right)\sqrt{2eB/\hbar
c}$\cite{ma:112102}. As $y_{0}$ approaches the sample edge, the
effect of the confining potential becomes important and it generates
the $k$-dependent dispersion of the energy levels, which has
profound effects on magnetotransport and magnetization properties.
Fig.\ref{energy} plots the the energy spectrum of the Hamiltonian
(\ref{h}) as a function of the cyclotron
center $y_{0}$.%

\section{The magnetization of the 2D holes\label{sec:OM_formula}}

Now we investigate the magnetization of the 2D holes in the III-V
semiconductor wells. According to the thermodynamics and the
statistical physics, one easily obtains that the magnetization
density is the derivative of the Helmholtz free energy density with
respect to $B$ at fixed electron density $\mathcal{N}$ and
temperature $T$, $M=-\left(\partial F/\partial
B\right)|_{\mathcal{N},T}$. For the present 2D hole model, the free
energy is given by \begin{align}
F(B,T) & =\mu\mathcal{N}-\frac{1}{L}\frac{N_{\nu}}{\beta}\int_{0}^{L}\mathrm{d}y_{0}\sum_{n,s}\ln\left\{ 1+e^{\beta\lbrack\mu-E_{n,s}(y_{0})]}\right\} \nonumber \\
 & \equiv\mu\mathcal{N}-\frac{1}{\beta}\int\mathrm{d}ED(E,B)\ln\{1+e^{\beta(\mu-E)}\},\label{Eq:FreeEnergy}\end{align}
 where $\beta$=$1/k_{B}T$, $N_{\nu}$=$eB/hc$, and $\mu$ is the
chemical potential. Note that we have defined in the above equation
the density of states (DOS) per area\begin{equation}
D(E,B)=\frac{N_{\nu}}{L}\sum_{n,s}\int_{0}^{L}\mathrm{d}y_{0}\delta(E-E_{n,s}(y_{0})).\label{DOS}\end{equation}
 The explicit inclusion of the DOS in the expression can be utilized
to take into account the impurity effect, which broadens the Landau
levels into Gaussian or Lorentzian in shape. For simplicity we do
not consider the broadening effect in this paper. In the absence of
edge states, the Landau Levels $E_{n,s}(y_{0})$ are uniform in space
and thus Eq. (\ref{Eq:FreeEnergy}) reduces to \begin{equation}
F(B,T)=\mu\mathcal{N}-\frac{N_{\nu}}{\beta}\sum_{n,s}\ln\left\{
1+e^{\beta(\mu-E_{n,s})}\right\} .\label{E2B}\end{equation}
 The $B$ dependent chemical potential $\mu$ is connected to the
experimentally accessible electron density $\mathcal{N}$ via the
local DOS. In the clean sample limit this is written as
\begin{equation}
\mathcal{N}=\frac{N_{\nu}}{L}\int_{0}^{L}\mathrm{d}y_{0}\sum_{n,s}f_{ns}\left(y_{0}\right),\label{E2C}\end{equation}
 where $f_{ns}\left(y_{0}\right)$=$\frac{1}{e^{\beta\left[E_{n,s}(y_{0})-\mu\right]}+1}$
is the Fermi distribution for the spin-split Landau levels
$E_{n,s}(y_{0})$. From Eq. (\ref{Eq:FreeEnergy}) the magnetization
density becomes\begin{align}
M & =\sum_{n,s}\left\{ -N_{\nu}\int_{0}^{L}\frac{\mathrm{d}y_{0}}{L}f_{ns}\left(y_{0}\right)\frac{\partial E_{n,s}(y_{0})}{\partial B}\right.\nonumber \\
 & \left.+\frac{e}{h}\frac{1}{\beta}\int_{0}^{L}\frac{\mathrm{d}y_{0}}{L}\ln\left\{ 1+e^{\beta\left[\mu-E_{n,s}(y_{0})\right]}\right\} \right\} \nonumber \\
 & \equiv M^{(0)}+M^{(1)}.\label{OM}\end{align}
 One can see that the magnetization consists of two parts. The first
part $M^{(0)}$ is the conventional contribution from the $B$
dependence of the Landau levels and thus is denoted as a
paramagnetic response. The second part $M^{(1)}$ comes from the $B$
dependence of the level degeneracy factor $N_{\nu}$, thus describing
the effect of the variation of the density of states upon the
magnetic field. Obviously, $M^{(0)}$ is negative while $M^{(1)}$ is
positive, the net result is an oscillation of the total
magnetization $M$ between the negative and positive values as a
function of $B$. At zero temperature, the expression for $M$ reduces
to a sum over all occupied Landau levels:
\begin{eqnarray}
M&=&\sum_{n,s}^{\text{occ}}\left\{
-N_{\nu}\int_{0}^{L}\frac{\mathrm{d}y_{0}}{L}\frac{\partial
E_{n,s}(y_{0})}{\partial B}\right.\nonumber\\
&&\left.+\frac{e}{h}\int_{0}^{L}\frac{\mathrm{d}y_{0}}{L}[\mu_{0}-E_{n,s}(y_{0})]\right\}
,\label{zeroOM}\end{eqnarray}
where the sum runs over all occupied
states and $\mu_{0}$ is the zero-temperature chemical potential
(Fermi energy).

\begin{figure}[ptb]
\includegraphics[scale=2.5]{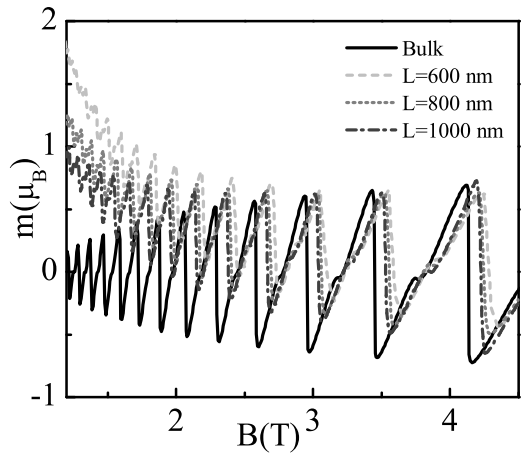}

\caption{The dHvA oscillations of the magnetization (per hole) of 2D
hole system with different system sizes. The parameters are chosen
as follows: the effective mass $m^{\ast}$=$0.27m_{e}$, the hole
number density $N_{h}$=$0.01$ nm$^{-2}$, the $g$-factor
$g_{s}$=$-0.44$, and the temperature $T=0.5$K. The Rashba SOC here
is neglected.}

\label{QSE}
\end{figure}

To clearly see the influence of the edge-states and Rashba SOC on
the magnetization, let us begin with the conventional result for the
bulk 2D holes without SOC and edge-state effects. In this case,  the
magnetization (per hole) $m$ (see the solid line in Fig. \ref{QSE})
displays the well-known sawtooth behavior with varying the magnetic
field $B$. One fact revealed in Fig. \ref{QSE} is that the inclusion
of the Zeeman splitting in the Landau levels will result in many
weak peaks appearing among the dHvA oscillation modes of the
physical quantities, the chemical potential $\mu$ and the
magnetization (per hole) $m$, at very low temperature (here
$T=0.2$K). These weak peaks have been observed recently by Schaapman
\emph{et al}\cite{PRB68.193308} when they measured the magnetization
of a dual-subband 2D electron gas, confined in a GaAs/AlGaAs
heterojection, and by Zhu \emph{et al}\cite{PRB67.155329} when they
measured the magnetization of high-mobility 2D electron gas. From
Fig. \ref{QSE} one obtain that these weak peaks can also be observed
in 2D hole system. These weak peaks will disappear when the
temperature turns sufficiently
high.%

\begin{figure}[ptb]
 \includegraphics[scale=2.]{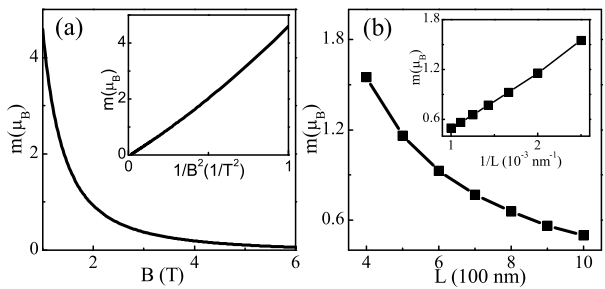}

\caption{(a) The $B$ dependence of the center of dHvA oscillations.
The system size $L$=$600$nm.\ (b) The dependence of the center of
dHvA oscillation on the size of the sample. The magnetic field $B$
is chosen around $2.0$T. In both figures the Rashba SOC is
neglected. The other parameters are same as those in Fig.
\ref{QSE}.}

\label{center}
\end{figure}

\begin{figure}[ptbptb]
 \includegraphics[scale=.8]{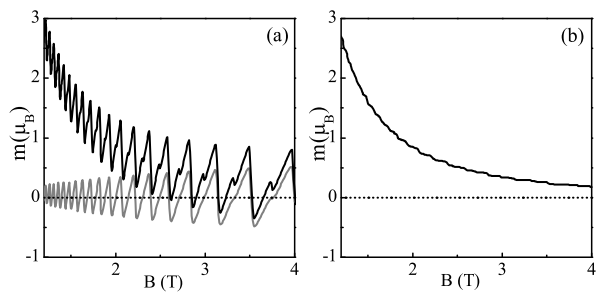}

\caption{(a) Gray curve: The bulk contribution of the magnetization
$m$ per hole (in units of $\mu_{B}$); Black curve: The total
magnetization per hole $m$ (in units of $\mu_{B}$). (b) The edge
contribution of the magnetization $m$ per hole (in units of
$\mu_{B}$). In both figures, the Rashba SOC $\alpha=0$ and
$L=600$nm. The other parameters are same as those in Fig.
\ref{QSE}.}

\label{part}
\end{figure}

Then we investigate the edge-state effects on the magnetization.
Fig. \ref{QSE} (the dashed, dotted, and dash-dotted lines) also
shows the influence of the edge states on the oscillations of
chemical potential and magnetization (dHvA oscillations) with
magnetic field. From Fig. \ref{QSE} one can easily observe a
prominent feature brought by the edge states, which is that the
center of the dHvA oscillations is now dependent on the magnetic
field. In particular, for the field less than $1$ Tesla, the
oscillatory magnetization is always positive in sign. Another
feature shown in Fig. \ref{QSE} is that the oscillation amplitude
decreases with decreasing the sample size. As is known, the origin
of dHvA oscillations is the degeneracy of Landau levels. The edge
states with dispersion lead to edge current, which not only is
crucial for the quantum Hall effects, but also very important for
the magnetization \cite{PRB59.7305}. The dispersion of the edge
states partially lift the degeneracy of the Landau levels. Thus the
edge states tend to destroy the dHvA oscillations. Therefore it
leads to the decreasing of oscillation amplitude as shown in Fig.
\ref{QSE}. The upshift of the center of dHvA oscillations may be
understood as follows: For the effects from the edge states, what
really matters is the ratio of two important length scales: the
magnetic length $l_{b}$ and the system size $L$. The decreasing of
$L$ is equivalent to the increasing of $l_{b}$, i.e., decreasing of
$B$ or $\omega_{c}$. From Eq. (\ref{OM}), one can see that with
decreasing of $L$, the second term overcomes the first term and
leads to the upshift of the center of dHvA oscillations. The smaller
the system size is, the more profound effects the edge states lead
to, as shown in Fig. \ref{QSE} for both the center and amplitude of
the dHvA oscillations. Fig. \ref{center}(b) shows quantitatively the
system size dependence of the shift of the oscillation center. It
has the dependence $1/L$. Roughly, the contribution of the edge
states is proportional to the number of edge states (as also seen
from Eq.(\ref{zeroOM})), which is proportional to $\nu r_{c}/L$,
where the cyclotron radius $r_{c}=\sqrt{\nu}l_{b}$
\cite{PhysRevB.70.235344}, with the number of the occupied Landau
levels $\nu\sim1/l_{b}^{2}$. Thus the center of dHvA oscillations is
proportional to $l_{b}^{4}/L=1/B^{2}L$. The $B$ and $L$ dependence
is clearly seen in Figs. \ref{center}(a) and (b). To see more
explicitly the contribution from edge states and bulk states, we
plot the total magnetization and the contribution from bulk
states in Fig. \ref{part}(a). 
The contribution from the edge states is obtained from Eq.
(\ref{zeroOM}) by summing over terms from edge states, with
$|y_{0}|<r_{c}$ or $|L-y_{0}|<r_{c}$. The rest contribution is from
bulk states. There is no upshift of the magnetization oscillation
center for the part from bulk states. It shows explicitly that the
upshift of the center of dHvA oscillations is due to the existence
of edge states. Fig. \ref{part}(b) shows the dependence of edge
states contribution on the magnetic field. The contribution from
edge states increases as decreasing the magnetic
field, or equivalently decreasing the sample size as one expects.%

\begin{figure}[ptb]
 \includegraphics[scale=0.7]{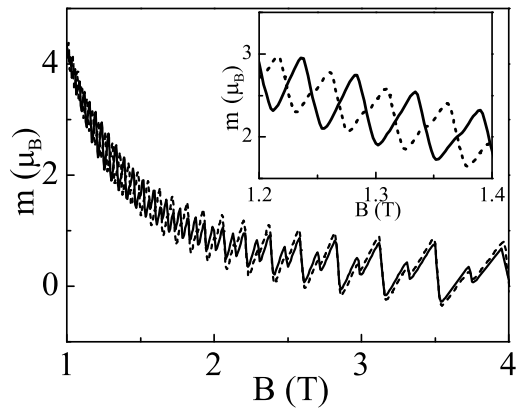}

\caption{The dHvA oscillations of the magnetization (per hole) of 2D
hole system as a function of external magnetic field $B$ with
(dashed line, the Rashba SOC strength $\alpha=20\,$meV\,nm$^{3}$)
and without (solid line) Rashba SOC coupling at $T=0.2$K$.$ Other
parameters are same as those in Fig. \ref{QSE}. The inset is the
enlarged magnetization oscillations between $1.2$T and $1.4$T. }

\label{Rashba}
\end{figure}

When the Rashba SOC is introduced, there is a energy competition
between the Zeeman coupling and the Rashba SOC. Also, due to the
entanglement between the orbital and spin degrees of freedom, it is
difficult to distinguish their separate contributions to the total
magnetization. These factors make the physical picture of the dHvA
oscillations to change fundamentally, as shown in Fig. \ref{Rashba}
for magnetization (per hole) $m$ as functions of $B$. One can see
from these two figures that the Rashba SOC has no visible influence
on the magnetic oscillations of the quantity  $m$ at large values of
$B$, where the Zeeman and spin-orbit coupling splitting are small
compared to the Landau level splitting. At low magnetic field,
however, the Rashba SOC modulation of the magnetic oscillations
becomes obvious, which can be clearly seen by the enlarged plots of
$\mu$ and $m$ in the inset in Fig. \ref{Rashba} for $B$ between
$1.2$T and $1.4$T. For comparison, we also re-plot in Fig.
\ref{Rashba} the cases without Rashba SOC. One can see from these
two figures that the SOC brings about two new features at low
magnetic field: (i) The sawtoothlike oscillating structure is
inversed, i.e., the location of peaks in $\mu$ and $m$ with SOC
correspond to the valleys without SOC. This inversion is due to the
different Landau levels in the two cases. (ii) The oscillation mode
is prominently modulated by SOC and a beating pattern appears. This
beating behavior in the oscillations are due to the fact that the
Landau levels are now unequally spaced due to
the presence of SOC.%

\begin{figure}[ptb]
 \includegraphics[scale=3]{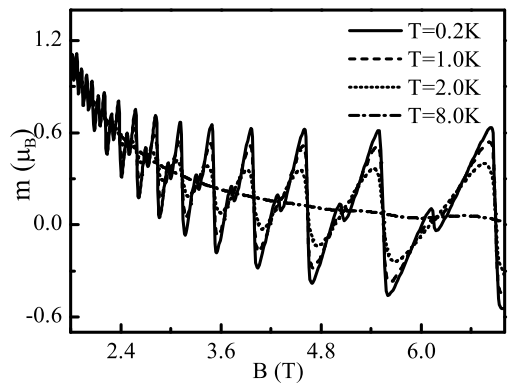}

\caption{The dHvA oscillations of the magnetization (per hole) of 2D
hole system as a function of the external magnetic field $B$ at
different temperatures. The Rashba SOC strength is set as
$\alpha=20\,$meV\,nm$^{3}$. The other parameters are same as those
in Fig. \ref{QSE}. }

\label{temp}
\end{figure}

Before ending this paper, let us briefly discuss the temperature
effect. It is well known that the infinite temperature will blur the
measured physical quantities, which including the dHvA oscillations
of the magnetization. To illustrate the temperature effect, we plot
in Fig.\ref{temp} the magnetizations of the holes at different
temperatures $T=0.5$, $1.0$, $2.0$, and $8.0\,$K, respectively.
Obviously with the temperature increases, the amplitude decreases
and the details of the oscillation tend to be blurred. At the
temperature $T=8.0$K, the dHvA oscillations vanish and all the
details are smeared out. However, the edge-state effect, which
depends on the sample
size and the external magnetic field, does not change with the temperature.%

\section{Summary}

In conclusion, we have theoretically investigated the dHvA
oscillations of the magnetization in the 2D heavy hole systems with
boundaries in a perpendicular magnetic filed. Especially, we focus
on the edge-state effect and the influence of the Rashba SOC on the
magnetization. The results show that the effect of the edge states
and Rashba SOC on the magnetization is less important when the
magnitude of the magnetic field $B$ is so large. However, with $B$
decreasing, the effect becomes evident. First, the dHvA oscillation
center of the magnetization will shift when the edge-state effect is
considered. The contribution of the edge states to the total
magnetization has a roughly linear relation with $1/B^{2}$ and
$1/L$. When the Rashba SOC introduced, the sawtoothlike oscillating
structure is inversed and a beating pattern appears. The phenomena
caused by the Rashba SOC can be blurred when the temperature turns
high. However, the edge-state effect do not change with the
temperature changing.

\subsection*{ACKNOWLEDGMENTS}

C. Fang and  S. S. Li were supported by the National Basic Research
Program of China (973 Program) grant No. G2009CB929300, and the
National Natural Science Foundation of China under Grant Nos.
60821061 and 60776061.   Z. G. Wang and P. Zhang were supported by
the National Natural Science Foundation of China under Grant Nos.
10604010 and 60776063.

\bibliographystyle{apsrev}
\bibliography{bibrefs}

\end{document}